\documentclass[aps,pre,twocolumn,showpacs,superscriptaddress]{revtex4}
\usepackage{epsfig}
\usepackage{times}
\begin{document}

\title{Facilitators on networks reveal the optimal interplay between information exchange and reciprocity}

\author{Attila Szolnoki}
\affiliation{Institute of Technical Physics and Materials Science, Research Centre for Natural Sciences, Hungarian Academy of Sciences, P.O. Box 49, H-1525 Budapest, Hungary}

\author{Matja{\v z} Perc}
\affiliation{Faculty of Natural Sciences and Mathematics, University of Maribor, Koro{\v s}ka cesta 160, SI-2000 Maribor, Slovenia}

\author{Mauro Mobilia}
\affiliation{Department of Applied Mathematics, School of Mathematics, University of Leeds, Leeds LS2 9JT, U.K.}

\begin{abstract}
Reciprocity is firmly established as an important mechanism that promotes cooperation. An efficient information exchange is likewise important, especially on structured populations, where interactions between players are limited. Motivated by these two facts, we explore the role of facilitators in social dilemmas on networks. Facilitators are here mirrors to their neighbors --- they cooperate with cooperators and defect with defectors --- but they do not participate in the exchange of strategies. As such, in addition to introducing direct reciprocity, they also obstruct information exchange. In well-mixed populations, facilitators favor the replacement and invasion of defection by cooperation as long as their number exceeds a critical value. In structured  populations, on the other hand, there exists a delicate balance between the benefits of reciprocity and the deterioration of information exchange. Extensive Monte Carlo simulations of social dilemmas on various interaction networks reveal that there exists an optimal interplay between reciprocity and information exchange, which sets in only when a small number of facilitators occupies the main hubs of the scale-free network. The drawbacks of missing cooperative hubs are more than compensated by reciprocity and, at the same time, the compromised information exchange is routed via the auxiliary hubs with only marginal losses in effectivity. These results indicate that it is not always optimal for the main hubs to become ``leaders of the masses'', but rather to exploit their highly connected state to promote tit-for-tat-like behavior.
\end{abstract}

\pacs{89.75.Fb, 87.23.Ge, 87.23.Kg}
\maketitle

\section{Introduction}
Unraveling the mechanisms at the origin of cooperation and understanding the reasons for so much biological diversity are among the most important challenges to Darwin's natural selection theory. For instance, it has been found that tropical forests and coral reefs teem with biological variation, and there are also many examples of insects that coordinate their efforts and even give up their own reproductive potential (fitness) to benefit that of the ``queen''~\cite{wilson_71}. Other examples include micro-organisms that can join forces to form biofilms and humans who are able to be ``supercooperators''~\cite{nadell_fems09, nowak_11}.

If only the fittest individuals survive and reproduce~\cite{dawkins_76}, why is there so
much diversity in nature~\cite{gaston_n00}? What are the mechanisms that originate and maintain
cooperative behavior? Evolutionary game theory (EGT) addresses these questions by means of simple but insightful models in which each individual's fitness varies and depends on the others' reproductive potential \cite{mestertong_01, nowak_06, sigmund_10}. EGT is the natural framework to mathematically study the dynamics of competing strategies (species), and the above fundamental questions have motivated a large body of work. In the context of EGT, understanding the evolution of cooperation often leads to a ``social dilemma'', such as in the  paradigmatic prisoner's dilemma game~\cite{axelrod_84}, where each rational individual chooses to defect (i.e. not to cooperate) while it would be in everyone's interest to cooperate. Cooperation dilemmas also arise in other EGT models such as the snowdrift and stag-hunt games \cite{hofbauer_98, szabo_pr07}.

Among the mechanisms that have been put forward to possibly explain the spread of cooperation, the influence of kin and group selection, as well as various forms of reciprocity (direct, indirect and network reciprocity), have been investigated, see e.g.~Refs.~\cite{hamilton_wd_jtb64a, trivers_qrb71, nowak_n92b, nowak_n98, hauert_n04, wang_z_epl12}. In particular, network reciprocity~\cite{szabo_pr07, roca_plr09, perc_bs10, perc_jrsi13}, whose principle has an appealing physical interpretation (cooperators are better off when they are surrounded by cooperators), has recently attracted interest in the physics community \cite{santos_prl05, chen_xj_pa07, gomez-gardenes_prl07, floria_pre09, chen_xj_pre09b, tanimoto_pre12, buesser_pre12, vukov_njp12, fu_srep12b, assaf_prl12, wang_z_pre12, wang_z_srep13c, fu_jsp13}. Quite interestingly it has been found that, in contrast to what happens in spatially-homogeneous (well-mixed) populations, the arrangement of individuals according to certain topologies can lead to very different scenarios. For instance, it was found that local interactions on regular lattices enhance the survival of cooperators in prisoner's dilemma games but inhibit their resistance against the invasion by ``defectors'' in snowdrift games~\cite{nowak_n92b, hauert_n04}.

Recently, the promotion of cooperation in the presence of cooperation facilitators has been investigated~\cite{mobilia_pre12, mobilia_csf13, mobilia_pre13}. These are special individuals who interact with competing players by mirroring their strategies, but they do not participate in the strategy exchange process. More precisely, they cooperate with cooperators and defect with defectors, but their status never changes over time, as they never adopt the strategy of another player. The influence of cooperation facilitators has been studied for the prisoner's dilemma, snowdrift and stag-hunt games in spatially-homogeneous populations. In such a setting, the mean field analysis and the cooperation fixation probability reveal that the invasion and replacement of defection by cooperation is favored when the number of facilitators exceeds a nontrivial critical value. When players are distributed on a structured population, however, we may face additional, competing effects. This is not only because each player has a limited interaction neighborhood, but also because facilitators, who do not participate in the strategy exchange process, can hinder the spread of information and so decelerate or even stop the invasion of the more successful strategy.

To clarify the impact of these effects, we consider evolutionary games where competing strategies and facilitators
are interpreted as species of a spatially-structured population. The fundamental question we aim to address is how the enhanced reciprocity on the one hand and the limited information exchange on the other hand interplay due to the presence of facilitators. To this end, we investigate the influence of facilitators (here, individuals facilitating either cooperators or defectors, see below) on a class of two-strategy games when individuals interact with their neighbors on a network. We specifically consider the cases of two-dimensional lattices and degree-homogeneous random graphs, as well as (heterogeneous) scale-free networks.

The organization of this paper is as follows: the models of social dilemmas with facilitators are introduced in the next section, and the main properties of the non-spatial prisoner's dilemma game with facilitators are outlined in Section~III. Numerical results for the level of cooperation in evolutionary games with facilitators on structured populations are presented and discussed in  Sections IV and V.

\section{Social dilemmas with facilitators}
We study pairwise evolutionary games on the square lattice, the random regular (degree-homogeneous) graph, and the Barab\'asi-Albert scale-free network \cite{barabasi_s99}, each with an average degree $k=4$ and size $N$.
Mutual cooperation yields the reward $R$, mutual defection leads to punishment $P$, and the mixed choice gives the cooperator the sucker's payoff $S$ and the defector the temptation $T$. Within this setup we have the prisoner's dilemma (PD) game if $T>R>P>S$, the snowdrift game (SG) if $T>R>S>P$, and the stag-hunt (SH) game if $R>T>P>S$, thus covering all three major social dilemma types. Without loss of generality, and for the sake of clarity, we set $R=1$, $P=0$, $0 \le T \le 2$, and $-1 \le S \le 1$, as illustrated in Fig.~\ref{ts}. We note that $T<1$ and $S>0$ quadrant marks the harmony game (HG), which however does not constitute a social dilemma. To further reduce the dimensionality of the parameter space, we introduce $T=1+r$ and $S=-r$, where $-1 \le r \le 1$ constitutes a diagonal across the $T-S$ plane that splits the harmony game and the prisoner's dilemma quadrant in half. Note that for $r<0$ we are in the harmony game quadrant, while for $r>0$ we are in the prisoner's dilemma quadrant. This parametrization of the prisoner's
dilemma game is the most challenging for the evolution of cooperation, and it is sometimes referred to as the donation game \cite{brede_pone13}.

Initially, in addition to the cooperators ($C$) and defectors ($D$) who are distributed uniformly at random in equal proportion, we designate a fraction $\rho_F$ of players as facilitators ($F$). Facilitators behave like mirrors to their neighbors, true to the most elementary form of reciprocity. A facilitator will cooperate with a cooperator, and it will defect with a defector. However, facilitators do not accumulate payoffs, and they do not participate in the exchange of strategies~\footnote{In Ref.~\cite{mobilia_csf13}, only facilitators cooperating with cooperators were considered.}. This means that facilitators can not be overtaken by other players, and they also can not spread. Accordingly, the fraction $\rho_F$ remains constant throughout the evolutionary process, and their positions on the network do not change. Within this setup, we seek to determine the optimal fraction of facilitators, as well as their impact on each particular social dilemma type.

We simulate the evolutionary process in accordance with the standard Monte Carlo simulation procedure comprising the following elementary steps. Among the subset of cooperators and defectors on the network, a randomly selected player $x$ acquires its payoff $P_x$ by playing the game with all its neighbors. Next, player $x$ randomly chooses one (also a non-facilitator) neighbor $y$, who then also acquires its payoff $P_y$ in the same way as previously player $x$. Lastly, player $x$ adopts the strategy $s_y$ from player $y$ with a probability determined by the Fermi function
\begin{equation}
\label{eq1}
W(s_y \to s_x)=\frac{1}{1+\exp[(P_x-P_y)/K]},
\end{equation}
where $K=0.1$ quantifies the uncertainty related to the strategy adoption process \cite{szabo_pr07}.
Note that $K$ can be interpreted as being proportional to the selection intensity, see e.g. \cite{nowak_06}.
In agreement with previous works, the selected value ensures that better-performing players are readily followed by their neighbors, although adopting the strategy of a player that performs worse is not impossible either \cite{perc_pre08b, szolnoki_njp08}.
This accounts for imperfect information, errors in the evaluation of the opponent, and similar unpredictable factors. We note however, that qualitatively identical behavior can be observed for other finite values of $K$ where the stochastic imitation dynamics remains non-neutral. Each full Monte Carlo step (MCS) gives a chance for every player to change its strategy once on average. All simulation results are obtained on networks with $N=10^4 - 2 \cdot 10^5$ players or more (including the facilitators) depending on the proximity to phase transition points, and the fraction of cooperators $\rho_C$ is determined in the stationary state after a sufficiently long relaxation (up to $2 \cdot 10^5$ MCS). To further improve accuracy, the final results are averaged over up to $100$ independent runs where interaction networks were generated 50 times for random and scale-free graphs at each set of parameter values.
\section{Non-spatial prisoner's dilemma with facilitators}
To better appreciate the influence of topology on social dilemmas in the presence of
facilitators, it is useful to outline the properties of the prisoner's dilemma
with facilitators  in the mean field setting and on a complete graph \cite{mobilia_csf13, mobilia_pre12}.
In this section, we focus on the prisoner's dilemma whose payoff matrix has entries $T$ for temptation
(with $1<T\leq 2$), $R=1$ for mutual defection, $P=0$ for punishment and $S$ (with $-1\leq S< 0$) as
Sucker's payoff, and assume $T+S\geq 1$.

In the mean field and complete graph settings,
 the population structure is homogeneous (``well-mixed'') and space therefore does not matter:
any individual can interact with all the others.
In a homogeneous population of size $N$, consisting of $j=N\rho_C$ cooperators, $k=N\rho_D$ defectors and
$\ell=N\rho_F$ facilitators,
the expected payoff of a cooperator is therefore
$\Pi^C_j=\frac{j+\ell-1}{N-1}+ S \frac{k}{N-1}$
and for a defector is
$\Pi^D_j=T\frac{j}{N-1}$
(self-interactions have been omitted). It is useful to introduce the payoff difference of competing strategies $\Delta \Pi_j=\Pi^D_j-\Pi^C_j$, as it is then easy to see that the difference consists of two terms $\Delta \Pi_j=\alpha (j/N)+\beta$, where the first cooperator dependent term contains $\alpha= \left(\frac{N}{N-1}\right)(T+S-1)$ while the fixed second term $\beta=\frac{1-S(N-\ell)-\ell }{N-1}$ depends only on the fraction of facilitators.

\subsection{The mean field limit}
The mean field limit (MF) limit corresponds to a spatially-homogeneous population of infinite size, $N\to \infty$.
In this situation, the dynamics of the prisoner's dilemma with facilitators
is described by a replicator-like equation for the density $\rho_C=j/N$ of cooperators
~\cite{taylor_p_mb78,schuster_jtb83,hofbauer_88,hofbauer_98,nowak_06,traulsen_09}. Here, since the underlying
dynamics is implemented with the Fermi rule~(\ref{eq1}),
such an equation reads~\cite{mobilia_csf13,mobilia_pre12}
\begin{eqnarray}
\label{MF}
\frac{d \rho_C}{d t}= -\rho_C(1-\rho_C-\rho_F)~\tanh{\left(\frac{\alpha \rho_C + \beta}{2K}\right)},
\end{eqnarray}
where in the MF limit, $\alpha=T+S-1\geq 0$ and $\beta=(S-1)\rho_F-S$.
The analysis of (\ref{MF}) readily reveals three distinct behaviors depending
on the fraction of facilitators $\rho_F$:
\begin{enumerate}
\item[(i)] When $\rho_F\leq S/(S-1)\equiv {\widetilde \rho_F}$, defection is still
the dominant strategy and the population evolves towards
$\rho_C=0$ and $\rho_D=1-\rho_F$ (only attractor).
\item[(ii)] On the other hand,
when ${\widetilde \rho_F}<\rho_F<1-T^{-1}$ and
$T+S>1$
the only attractor of  (\ref{MF})
is $\rho_{C}^*=-\beta/\alpha=\frac{S+(1-S)\rho_F}{T+S-1}$. There is a stable  coexistence
of cooperators and defectors.\\
\item[(iii)] When $\rho_F>1-T^{-1}$ and $T+S>1$, cooperation is
the dominant strategy and the dynamics approaches
$\rho_C=1-\rho_F$ and $\rho_D=0$.
\end{enumerate}
It is worth noting that Eq.~(\ref{MF}) has no coexistence steady state
when $T+S=1$ (since $\alpha=0$).
The MF dynamics along such a special line reproduces the behaviors (i) and (iii):
 $\rho_C=1-\rho_F$ is stable when $\rho_F>{\widetilde \rho_F}$ (since $\beta>0$) and unstable
otherwise, with $\rho_C=0$ being the only attractor when
 $\rho_F<{\widetilde \rho_F}$.

\subsection{The case of complete graphs ($N<\infty$)}
When the population is well-mixed and of finite size, $N<\infty$, its
evolution is usually described in terms of a birth-and-death Markov
chain with absorbing boundaries~\cite{nowak_06,traulsen_09,mobilia_csf13,mobilia_pre12}.
In this case, the {\it fixation} of either defection ($\rho_C=0$) or cooperation ($\rho_C=1-\rho_F$)
is guaranteed.
On complete graphs, the dynamics is implemented as a Markov chain
 with rates $T_{j}^{\pm}=\frac{j(N-\ell-j)}{N(N-1)}~\left[1+e^{\pm (\alpha j + N \beta)/NK}\right]^{-1}$
 for the transitions $j \to j \pm 1$.

Since fluctuations prevent stable coexistence when $N<\infty$, it is important to understand
when cooperation is favored by
selection. The following conditions have been proposed~\cite{nowak_n04b,nowak_06}: (1) the invasion by cooperators is favored
when $\Delta \Pi_1<0$; (2) selection favors the replacement of defection by cooperation
when $(N-\ell)\phi_C >1$, where
$\phi_C=\left[1+\sum_{n=1}^{N-\ell -1} {\rm exp}\left(\frac{n}{2NK}[\alpha(n+1)+2N\beta]\right)\right]^{-1}$
is the fixation probability of a single cooperator~\cite{mobilia_csf13}.

When $N\gg \ell$, the invasion condition (1) is satisfied when $\rho_F>{\widetilde \rho_F}$,
while the replacement condition (2) reads
\begin{eqnarray}
\label{eq3}
N-\ell>\sum_{n=1}^{N-\ell}  {\rm exp}\left[
\frac{\alpha n}{K}\left(\frac{n}{2N}+\frac{\beta}{\alpha}
\right)
\right]
\end{eqnarray}
and is satisfied when $\rho_F>\rho_F^{*}$, where $\rho_F^{*}$
is a critical value obtained by equating both sides of (\ref{eq3}). It has been found that
$\rho_F^{*}\geq {\widetilde \rho_F}$ when $T+S>1$
and $\rho_F^{*}\leq {\widetilde \rho_F}$ otherwise (with $\rho_F^{*}= {\widetilde \rho_F}$ when $T+S=1$)~\cite{mobilia_csf13}.

In summary, in the MF limit cooperators and defectors coexist when the fraction of facilitator $\rho_F$
exceeds the critical value ${\widetilde \rho_F}=S/(S-1)$ and $T+S \geq 1$, whereas
cooperation is favored on complete graphs when $\rho_F$ is above
a critical value  $\rho_F^{*}\geq {\widetilde \rho_F}$.

\section{Results on networks}
\begin{figure}
\centerline{\epsfig{file=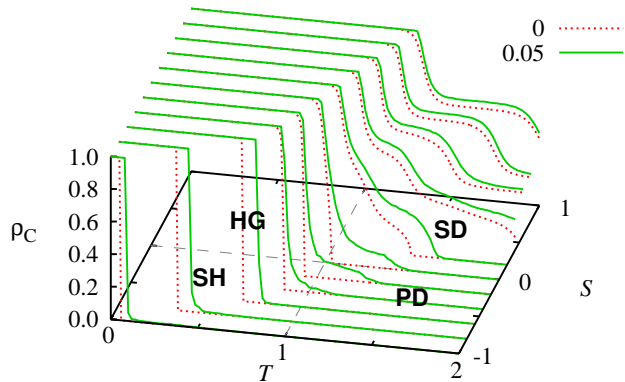,width=8.5cm}}
\caption{(Color online) Evolution of cooperation with and without facilitators on the square lattice. Depicted is the rescaled stationary fraction of cooperators $\rho_C$ on the whole $T-S$ parameter plane, as obtained in the absence of facilitators (red dotted lines) and with $\rho_F=0.05$ (green solid lines). It can be observed that facilitators do not change the qualitative properties of the solutions, but their presence does shift the survival barrier of cooperators towards harsher conditions, especially in the prisoner's dilemma quadrant, see text.}
\label{ts}
\end{figure}

We begin by studying the impact of facilitators on the square lattice with periodic boundary conditions. The results are summarized in Fig.~\ref{ts}. For a comprehensive insight, we compare the outcomes of the evolutionary process on the whole $T-S$ parameter plane, as obtained with and without facilitators. To allow for a better comparison of the influence of
facilitators, in Fig~\ref{ts} and in the other figures, we report the relative density of cooperators obtained by rescaling the physical fraction of cooperators present in the population by $(1-\rho_F)^{-1}$,  i.e. we have rescaled $\rho_C\to \rho_C/(1-\rho_F)$ so that in all the figures its value always ranges between $0$ and $1$.
The presented results indicate that the impact of facilitators can be considered as a second-order effect. While the results do not change qualitatively, the survival threshold of cooperators shifts considerably towards harsher conditions. This is most pronounced in the prisoner's dilemma (PD) quadrant, although quantitative changes are observable in the snowdrift (SD) and the stag-hunt (SH) quadrant as well. Facilitators exercise a second-order effect because the outcome is primarily determined by the fact that the evolutionary games are staged on a structured population (in this case the square lattice). The spatiality of interactions always allows cooperators and defectors to coexist in a special parameter interval while the presence of facilitators shifts the borders of different stable solutions. This behavior is significantly different form the non-spatial behavior of evolutionary games with facilitators outlined in the previous section, where their presence can radically change the character of solutions and the type of the social dilemma. In Fig.~\ref{ts} we also notice that nothing uncharacteristic happens along the line $T+S=1$. Henceforth, we will characterize the comprehensive properties of the evolutionary games on networks  by conveniently focusing on the parametrization $T=1+r, S=-r$, with $-1\leq r\leq 1$. This parametrization constitutes a diagonal across the prisoner's dilemma and harmony game quadrant.
\begin{figure}
\centerline{\epsfig{file=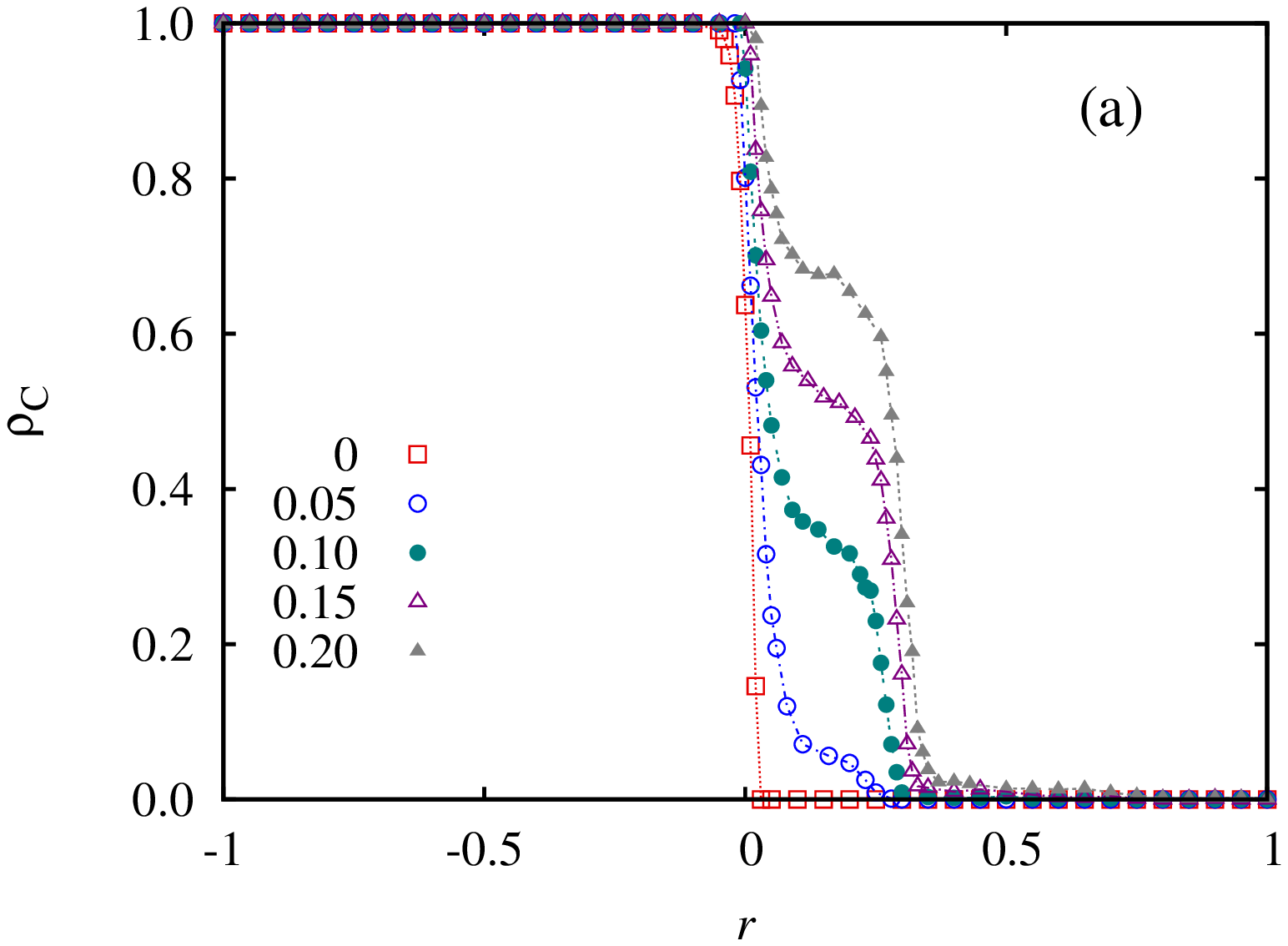,width=8cm}}
\centerline{\epsfig{file=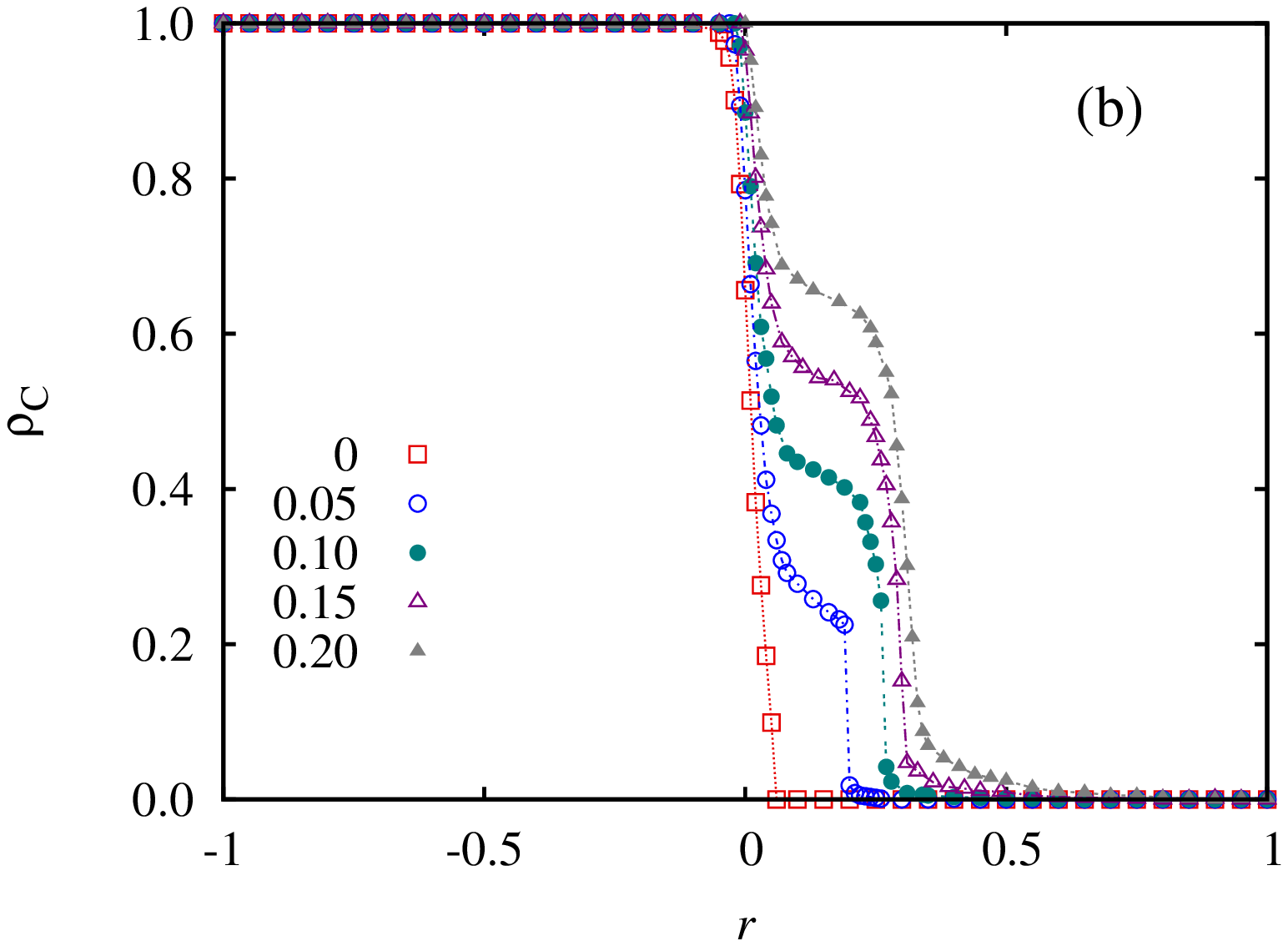,width=8cm}}
\caption{The impact of facilitators on the square lattice (top) and the random regular graph (bottom). Depicted is the stationary fraction of cooperators $\rho_C$ in dependence on $r$, as obtained for different fractions of facilitators occupying the network (see figure legend for the values of $\rho_F$). Due to the qualitatively identical results obtained on the two networks, it can be concluded that the topology of the interaction network does not play a notable role. More precisely, if the network remains degree-homogeneous, then the randomness of interactions yields the same results as lattice-type models.}
\label{sqr_rrg}
\end{figure}

\begin{figure}
\centerline{\epsfig{file=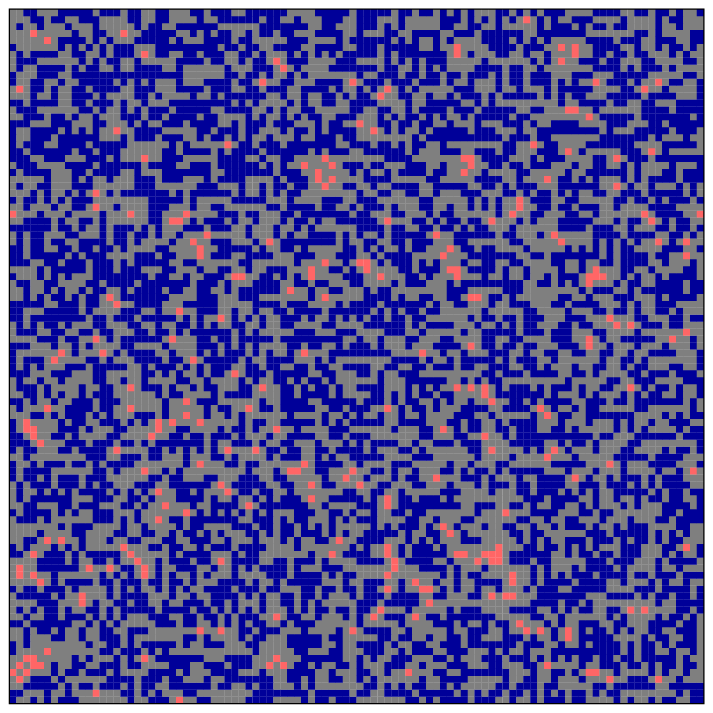,width=4.25cm}\epsfig{file=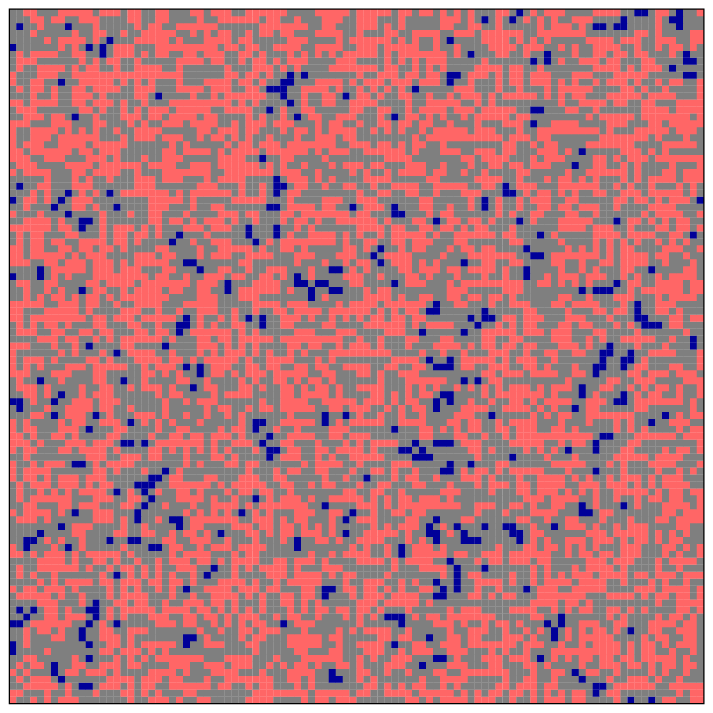,width=4.25cm}}
\caption{(Color online) Characteristic distributions of cooperators (blue, dark grey) and defectors (red, light grey) on the $L\times L= 100 \times 100$ square lattice, as obtained for $r=-1$ (left) and $r=1$ (right) if the fraction of facilitators (gray) is sufficiently high for them to split the population in effectively isolated smaller fragments. The left panel depicts the outcome of the most lenient harmony game, yet still some defectors are able to survive. On the contrary, the right panel depicts the outcome of the harshest prisoner's dilemma game, yet cooperators survive. In both  panels the fraction of facilitators is $\rho_F=0.5$.}
\label{snaps}
\end{figure}

Next, we explore how the topology of the interaction network affects the impact of facilitators. To avoid effects stemming from the heterogeneity of the interaction network, we first compare the outcomes obtained on the square lattice and the random regular (degree-homogeneous) graph. On both these networks every player has four neighbors ($k=4$). As Fig.~\ref{sqr_rrg} shows, the principal impact of facilitators is to widen the parameter range where $C$ and $D$ players coexist. Moreover, increasing $\rho_F$ increases the fraction of cooperators within this interval, and as expected, contribute to a higher level of cooperation in the population. However, if the fraction of facilitators becomes too high, typically $\rho_F>0.4$, then facilitators will no longer play solely the role of mirrors to their neighbors, but they will also serve as ``walls'' that prevent efficient information spreading throughout the system. At this point, it is worth reiterating that facilitators do not participate actively in the evolutionary process. Consequently, too many facilitators will separate competing strategies, and there will be segregation with the population splitting apart into smaller fragments. Within these small and effectively isolated regions, the parametrization of the game, and thus the type of the social dilemma, no longer plays a decisive role for the survival of the two competing strategies. Effectively, a ``dilemma hiding'' effect sets in, where the
prevailing configuration is determined the local initial conditions and remains frozen afterwards.
This means that, after a very short initial period, the strategies can no longer evolve according to the dynamics that would be dictated by the payoff elements. The ultimate consequence of the ``dilemma hiding'' effect is that, within the locally frozen states, some cooperators may survive even at the most demanding conditions that constitute a prisoner's dilemma $(r=1)$, and vice versa, some defectors may survive even at the most lenient conditions that characterize the harmony game $(r=-1)$. Two representative snapshots depicting such an evolutionary outcome are presented in Fig.~\ref{snaps}.

\begin{figure}
\centerline{\epsfig{file=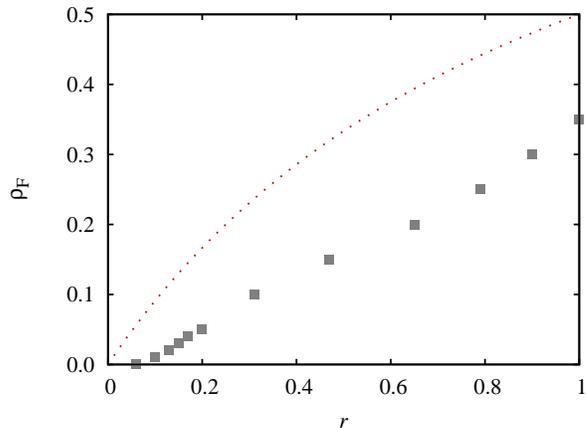,width=8cm}}
\caption{The minimal fraction of facilitators that is necessary to avoid the tragedy of the commons (the pure $D$ state) as a function of $r=T-1$ for the prisoner's dilemma: Cooperation becomes viable above this threshold. Comparison of the mean field prediction ${\widetilde \rho_F}=r/(1+r)$ (dotted line) with the value obtained by numerical simulations on random  graphs with regular degree $4$ (symbols). When the degree increases, the symbols would move towards the mean field line (not shown here). See also main text for details.}
\label{rhoFcrit}
\end{figure}

It is also worth comparing the results obtained on random graphs with degree $k=4$ and the predictions obtained for complete graphs where the degree is equal to $N$: on the latter, at a fixed value of $\rho_F$, we have shown that cooperation prevails when $r<\rho_F/(1-\rho_F)$. On the other hand, if the node degree is four then $\rho_C\approx 1$ when $r \le 0$ while $\rho_C\approx 0$ when $r \gtrsim 3 \rho_F$. The comparison of critical facilitator density with the results of numerical simulations for the random regular graphs with $k=4$ in Fig.~\ref{rhoFcrit} reveals that the critical threshold on the latter is always below the mean field prediction ${\widetilde \rho_F}$. This indicates that less facilitators are needed on a random regular graph with a finite degree than on a complete graph for cooperation to prevail.

\begin{figure}[ht!]
\centerline{\epsfig{file=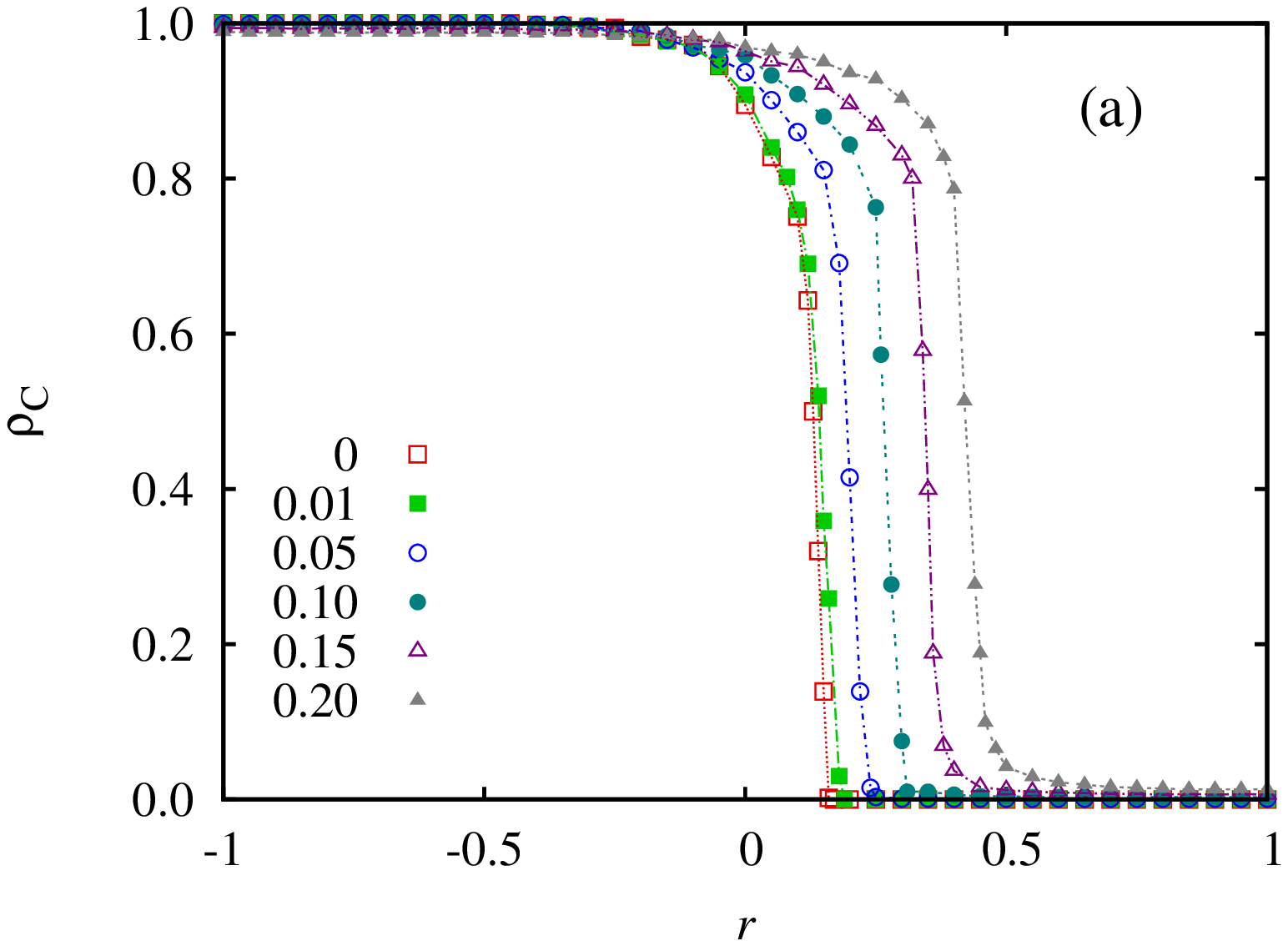,width=7.7cm}}
\centerline{\epsfig{file=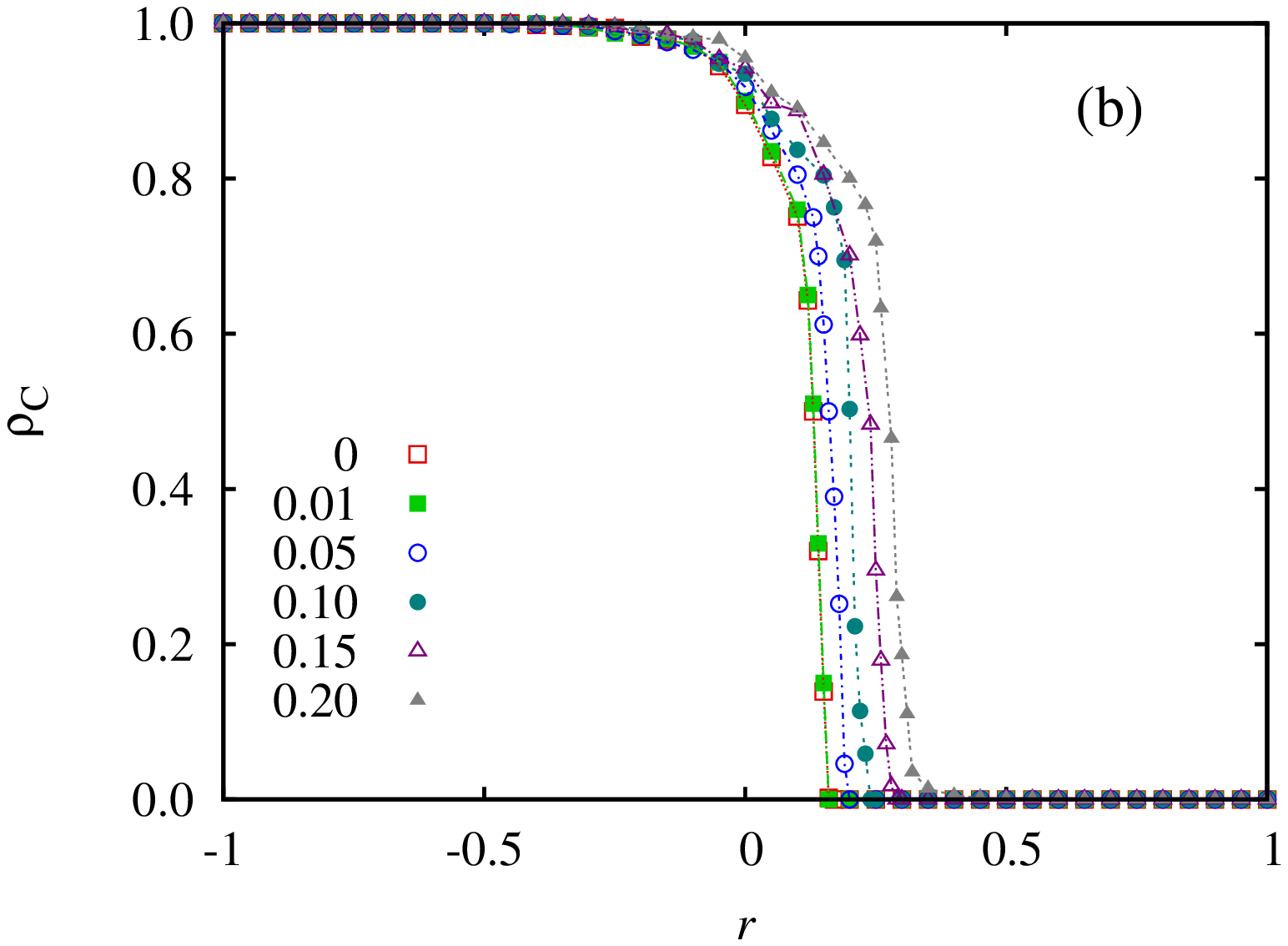,width=7.7cm}}
\centerline{\epsfig{file=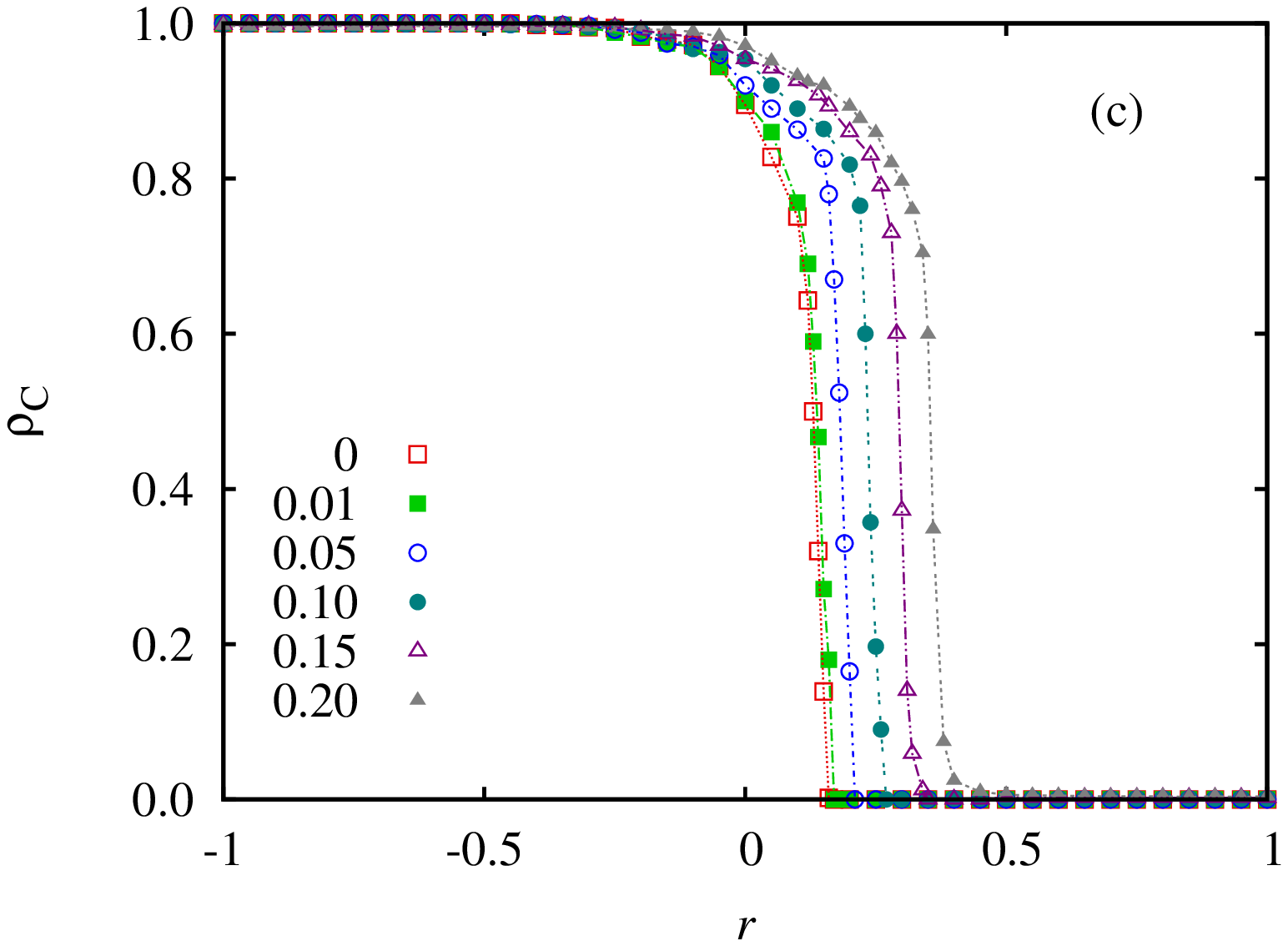,width=7.7cm}}
\caption{The impact of facilitators on the scale-free network if their placement is uniformly random regardless of the degree of players (top), or if their placement is limited to players with low (middle) or intermediate (bottom) degree. Depicted is the stationary fraction of cooperators $\rho_C$ in dependence on $r$, as obtained for different fractions of facilitators occupying the network (see figure legend for the values of $\rho_F$). As in Fig.~\ref{sqr_rrg}, increasing the value of $\rho_F$ will significantly extend the region where cooperators and defectors are able to coexist, especially if the facilitators are placed randomly (top). The results are obtained using $N=10^5$ system size, see text.}
\label{sf_rsm}
\end{figure}

So far, we have considered only homogeneous interaction networks, where the distribution of facilitators was always uniformly random, and the specific placement did not matter because all players on the square lattice and the random regular graph have the same degree. This changes if instead we apply heterogeneous interaction graphs, like the scale-free networks, where the distribution of degree is a power law. We use the algorithm proposed by Barab{\'a}si and Albert \cite{barabasi_s99} to construct scale-free networks with the average degree $k=4$ and degree distribution $P_k\sim k^{-3}$ (BA graphs), and we consider four different cases of where on the network to place facilitators. First, to keep the analogy with the previous treatment on homogeneous networks, we choose players uniformly at random regardless of their degree. As results presented in Fig.~\ref{sf_rsm} (top) illustrate, increasing $\rho_F$ will not just increase $\rho_C$, but it will also expand gradually the coexistence region significantly toward stronger social dilemmas (higher values of $r$).

Naturally, we could also observe the ``dilemma hiding'' effect for sufficiently high values of $\rho_F$ (now shown), which for the considered scale-free network and randomly distributed facilitators begins at $\rho_F \approx 0.5$. If, on the other hand, facilitators are placed on low or intermediate degree nodes, the ``dilemma hiding'' effect appears only at even larger values of $\rho_F$. This is understandable since low degree nodes have a lower number of links to the other players, and hence disabling their ability to transfer information obviously has a lesser impact than if one of the network hubs would loose this ability. In terms of the impact of facilitators on the evolution of cooperation, however, placing facilitators on low or intermediate degree nodes has qualitatively the same impact as placing them randomly across the whole network. As evidenced by the results presented in Fig.~\ref{sf_rsm} (middle and bottom), the only difference is that the shift of the border where both strategies can coexist is obviously smaller if facilitators occupy low degree nodes, and it is slightly larger if facilitators occupy intermediate degree nodes. The shift is the largest if the placement of facilitators is uniformly random regardless of the degree of players, presumably because some facilitators then also occupy the hubs of the network.

\begin{figure}
\centerline{\epsfig{file=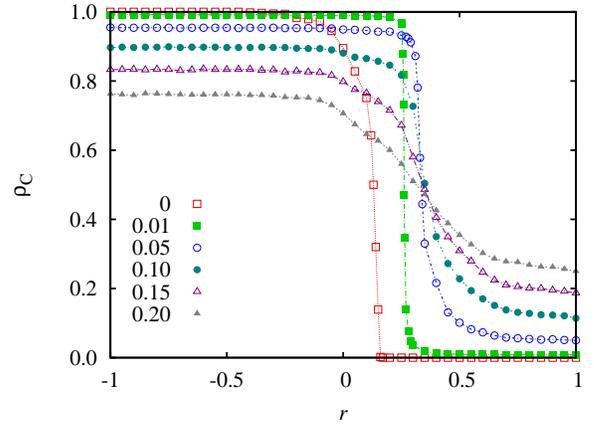,width=7.9cm}}
\caption{The impact of facilitators on the BA scale-free network if their placement is limited to players with high degree. Depicted is the stationary fraction of cooperators $\rho_C$ in dependence on $r$, as obtained for different fractions of facilitators occupying the targeted high degree nodes (see figure legend for the values of $\rho_F$ placed at the most connected nodes). These results reveal the existence of the optimal interplay between information exchange and reciprocity (see main text for details). Compared to the results presented in Fig.~\ref{sf_rsm}, only in this particular case is it possible to combine the two effects to arrive at the best conditions for widespread cooperation. The results are obtained using $N=10^5$ system size.}
\label{sf_large}
\end{figure}

Studying the impact of facilitators targeted on high degree nodes will resolve this ambiguity, but before presenting the results, it is worth emphasizing that the expectations are rather conflicting for this particular case. On the one hand, we may hope that placing the facilitators on the hubs will improve the cooperation level even further because their special status can enhance network reciprocity (this hope is also justified by the preceding results presented in Fig.~\ref{sf_rsm}).
On the other hand, it is precisely this special position of facilitators that brings this expectation into questioning. As demonstrated in several previous works \cite{santos_prl05, szabo_pr07, szolnoki_pa08}, hubs of scale-free networks play a crucial role in ensuring highly cooperative states under adverse conditions. Only the cooperative hubs can reap long-term benefits from their highly connected status, and thus serve as a lucid reminder of the benefits of cooperative behavior. However, if we place facilitators on the hubs, then this mechanism can no longer work. Effectively, we remove the cooperative leaders and replace them with ``mirrors'' instead. We emphasize again that here facilitators cannot be ``followed'', i.e., they just exactly reciprocate the strategy of each of their neighbors. Consequently, the level of cooperation may drop back to the level we observe on homogeneous networks. Another drawback of placing facilitators on the hubs is the hindering of the information flow through the system, which in this case is particularly effective and can thus easily evoke the ``dilemma hiding'' effect demonstrated in Fig.~\ref{snaps}.

All these arguments make the results presented in Fig.~\ref{sf_large}, which were obtained by placing facilitators on the high degree nodes of the BA scale-free network, especially interesting. These results partly fulfill our expectations outlined in the previous paragraph, but there are also some unexpected outcomes. More precisely, the ``dilemma hiding'' effect emerges at rather small $\rho_F$ values. If the top 5\% of nodes are occupied by facilitators, for instance, then we can observe cooperators surviving even at the highest $r$ value, but some defectors prevail in the harmony game region ($r<0$) as well, thus indicating an imperfect information flow. This effect is even more evident at higher densities of facilitators. However, if  only the top 1\% of nodes host facilitators, then the information exchange remains practically flawless, but at the same time a significant improvement in cooperation level due to active reciprocity can be observed too. Here the region of near complete cooperation dominance is extended toward significantly higher $r$ values (shifted from $r_c \approx -0.15$ to $r_c \approx +0.2$), which is surprising because players cannot imitate the main hubs. Still, some prominently-placed facilitators (mirrors) are able not just to compensate the impaired learning process, but even promote cooperation more efficiently then a flawlessly learning process would do. Naturally, in this case too the spreading of cooperative behavior happens predominantly via learning, but not through the most obvious channels --- via the strongest hubs --- but rather via the slightly less dominant nodes of the scale-free networks. More precisely, indirect connections between less preferred players around the hubs still work, which enables the spreading of the most successful strategy. At the same time, the advantages of cooperation are massively amplified by facilitators, which introduce direct reciprocity that pays more than undisturbed learning. We note that the results presented here for BA graphs are expected to hold for scale-free networks of degree-distribution $P_k\sim k^{-\gamma}$ with $1<\gamma\leq 3$ that are characterized by high-degree nodes, while we expect to recover the random degree-homogeneous scenario when $\gamma>3$ ("hubs" are then unlikely).

\section{Discussion}
We have studied the role of facilitators on structured populations. Facilitators are the ideal mirror to their neighbors, and as such they introduce reciprocity directly to the studied evolutionary games. Results obtained for well-mixed populations show that facilitators favor the evolution of cooperation as long as they are sufficiently present in the population. Importantly, there are no negative consequences even if their numbers become large. On structured populations this no longer holds, because in addition to reciprocity, facilitators also obstruct information exchange. Here, facilitators cooperate with cooperators and defect with defectors, but they do not participate in the exchange of strategies, meaning that they can not be overtaken by other players, and they also can not spread. Accordingly, we have shown that if the facilitators are too many, they no longer play solely the role of mirrors to their neighbors, but they also act as ``walls'' that prevent efficient information spreading throughout the system. These walls separate competing strategies, and they compartmentalize the population into effectively isolated regions. Within these regions the type of the social dilemma no longer plays a decisive role for the survival of the two competing strategies, and effectively a ``dilemma hiding'' effect sets in. Only if the fraction of facilitators is sufficiently small is the evolution of cooperation promoted, in particular by extending the survival region of cooperators towards harsher conditions. Besides homogeneous networks such as the square lattice and the random regular graph, we have also considered heterogeneous interaction networks --- the most representative being the Barab{\'a}si-Albert scale-free network --- where the placement of facilitators plays a decisive role. If the facilitators occupied the main hubs of the network, we were able to observe the optimal interplay between the benefits of reciprocity and the drawbacks of hindered information exchange. This result is highly counterintuitive because previous research has strongly emphasized the crucial role of cooperative hubs for the successful evolution of cooperation \cite{santos_prl05, szabo_pr07, szolnoki_pa08}. According to established previous reasoning, hubs are able influence their large neighborhoods directly, which yields large homogeneous domains and thus facilitates the manifestation of long-term benefits of cooperation. Here we have found that hubs can work even better in favor of cooperative behavior if they are not used as ``leaders of the masses'', but rather as mirrors to their many neighbors. As an avenue to explore in the future, it could be interesting to study how the results on heterogeneous graphs might change if we apply different degree distributions of nodes. If we decrease the number of hubs, for example, then the results may tend towards those we have obtained on regular graphs where there are no distinguished players.

Summarizing, we have shown that reciprocity outperforms imitation via learning, and that the latter can still be effectively enough re-routed through the auxiliary hubs. This delicate balance between augmented reciprocity and information exchange proves to be the best combination that is able to maintain cooperation even at the most adverse conditions, while at the same time disallowing widespread defection at lenient conditions. Interestingly, it could be better to promote tit-for-tat-like behavior in prominent players rather than for them to aspire towards leader-follower relations.

\begin{acknowledgments}
This research was supported by the Hungarian National Research Fund (Grant K-101490), TAMOP-4.2.2.A-11/1/KONV-2012-0051, and the Slovenian Research Agency (Grant J1-4055). We would also like to thank the referees for their insightful comments.
\end{acknowledgments}

\end{document}